\documentstyle[12pt]{article}

\makeatletter
\def\bigans{y }
\read-1 to\answ
\ifx\answ\bigans

\ifcase\@ptsize
 \font\tenmsa=msam10
 \font\sevenmsa=msam7
 \font\fivemsa=msam5
 \font\tenmsb=msbm10
 \font\sevenmsb=msbm7
 \font\fivemsb=msbm5
\or
 \font\tenmsa=msam10 scaled \magstephalf
 \font\sevenmsa=msam8
 \font\fivemsa=msam6
 \font\tenmsb=msbm10 scaled \magstephalf
 \font\sevenmsb=msbm8
 \font\fivemsb=msbm6
\or
 \font\tenmsa=msam10 scaled \magstep1
 \font\sevenmsa=msam8
 \font\fivemsa=msam6
 \font\tenmsb=msbm10 scaled \magstep1
 \font\sevenmsb=msbm8
 \font\fivemsb=msbm6
\fi

\newfam\msafam
\newfam\msbfam
\textfont\msafam=\tenmsa  \scriptfont\msafam=\sevenmsa
  \scriptscriptfont\msafam=\fivemsa
\textfont\msbfam=\tenmsb  \scriptfont\msbfam=\sevenmsb
  \scriptscriptfont\msbfam=\fivemsb

\def\hexnumber@#1{\ifnum#1<10 \number#1\else
 \ifnum#1=10 A\else\ifnum#1=11 B\else\ifnum#1=12 C\else
 \ifnum#1=13 D\else\ifnum#1=14 E\else\ifnum#1=15 F\fi\fi\fi\fi\fi\fi\fi}

\def\msa@{\hexnumber@\msafam}
\def\msb@{\hexnumber@\msbfam}

\catcode`\@=12

\def\CC{\Bbb C}

\else
\def\CC{I\!\!\!\!C}

		\fi

\catcode`@=11
\def\citen#1{\if@filesw \immediate\write \@auxout {\string\citation{#1}}\fi%
\@tempcntb\m@ne \let\@h@ld\relax \def\@citea{}%
\@for \@citeb:=#1\do {\@ifundefined {b@\@citeb}%
    {\@h@ld\@citea\@tempcntb\m@ne{\bf ?}%
    \@warning {Citation `\@citeb ' on page \thepage \space undefined}}%
    {\@tempcnta\@tempcntb \advance\@tempcnta\@ne
    \setbox\z@\hbox\bgroup\ifcat0\csname b@\@citeb \endcsname \relax
    \egroup \@tempcntb\number\csname b@\@citeb \endcsname \relax
    \else \egroup \@tempcntb\m@ne \fi \ifnum\@tempcnta=\@tempcntb
    \ifx\@h@ld\relax \edef \@h@ld{\@citea\csname b@\@citeb\endcsname}%
    \else \edef\@h@ld{\hbox{--}\penalty\@highpenalty
    \csname b@\@citeb\endcsname}\fi
    \else \@h@ld\@citea\csname b@\@citeb \endcsname \let\@h@ld\relax \fi}%
\def\@citea{,\penalty\@highpenalty\hskip.13em plus.13em minus.13em}}\@h@ld}
\def\@citex[#1]#2{\@cite{\citen{#2}}{#1}}%
\def\@cite#1#2{\leavevmode\unskip\ifnum\lastpenalty=\z@\penalty\@highpenalty\fi%
   $^{\scriptscriptstyle \multiply\@highpenalty 3 \mbox{\rm\scriptsize#1%
  \if@tempswa,\penalty\@highpenalty\ #2\fi}}$}   %

\def\dcite{\@ifnextchar [{\@tempswatrue\@dcitex}{\@tempswafalse\@dcitex[]}}

\def\@dcitex[#1]#2{\if@filesw\immediate\write\@auxout{\string\citation{#2}}\fi
  \def\@dcitea{}\@dcite{\@for\@dciteb:=#2\do
    {\@dcitea\def\@dcitea{,}\@ifundefined
       {b@\@dciteb}{{\bf ?}\@warning
       {d(line)cite  `\@dciteb' on page \thepage \space undefined}}%
\hbox{\csname b@\@dciteb\endcsname}}}{#1}}

\def\@dcite#1#2{$\mbox{\rm#1\if@tempswa , #2\fi}$}
\catcode`@=12

\def\onward{\addtocounter{section}{1} \setcounter{equation}{0} }

\def\smultab#1#2#3#4#5#6#7{\put (0,4){\line(1,0){#1}}
                    \multiput(0,3)(1,0){#1}{\line(1,0){1}}
                    \multiput(1,3)(1,0){#1}{\line(0,1){1}}
                    \multiput(0,2)(1,0){#2}{\line(1,0){1}}
                    \multiput(1,2)(1,0){#2}{\line(0,1){1}}
                    \multiput(0,1)(1,0){#3}{\line(1,0){1}}
                    \multiput(1,1)(1,0){#3}{\line(0,1){1}}
                    \multiput(0,0)(1,0){#4}{\line(1,0){1}}
                    \multiput(1,0)(1,0){#4}{\line(0,1){1}}
                    \multiput(0,-1)(1,0){#5}{\line(1,0){1}}
                    \multiput(1,-1)(1,0){#5}{\line(0,1){1}}
                    \multiput(0,-2)(1,0){#6}{\line(1,0){1}}
                    \multiput(1,-2)(1,0){#6}{\line(0,1){1}}
                          \put (0,4){\line(0,-1){#7}}}
\def\young#1#2#3#4#5#6#7{\begin{picture}(#1,#7)(0,3)
                \thicklines \smultab#1#2#3#4#5#6#7 \end{picture}}

\def\one{{\setlength{\unitlength}{0.15cm}\young1000001}}
\def\bone{{\setlength{\unitlength}{0.20cm}\young1000001}}
\def\two{{\young2000001}}
\def\btwo{{\setlength{\unitlength}{0.20cm}\young2000001}}
\def\oneone{{\raise 1mm \hbox{\young1100002}}}
\def\boneone{{\raise 1mm
                \hbox{{\setlength{\unitlength}{0.20cm}\young1100002}}}}
\def\bfid{{\bf 1}}
\def\Tr{{\rm Tr}}
\def\sun{{\rm SU}(N)}
\def\spn{{\rm Sp}(N)}
\def\son{{\rm SO}(N)}
\def\nbr{i}
\def\ncc{k}
\def\kap{\kappa}
\def\cS{{\cal S}}
\def\cK{{\cal K}}
\def\cZ{{\cal Z}}
\def\cA{{\cal A}}
\def\cW{{\cal W}}
\def\sign{\eta}
\def\Eu{{\cal E} }
\def\setU{ { U_1,\ldots, U_b } }
\def\setR{ { R_1, \ldots, R_b } }
\def\sameR{ {R,\ldots,R} }
\def\setkap{ { \kap_1, \ldots, \kap_b  } }
\def\NG#1#2#3{N_G(#1,#2;#3)}
\def\Up{\Upsilon^{G}}
\def\Upkap{ \Up_{\kap} }
\def\gchi{  \hchi^{G} }
\def\Cover{ {\cal C}}
\def\Covers{  \Cover_\rho(\cS, r, \setkap,  \nbr, k) }
\def\Omr{  \Omega_r }
\def\Om{\Omega}
\def\kinK{  \kap \in {\cal K}_r}
\def\RinY{  R \in Y_r}
\def\col{k}
\def\smb{{\footnotesize \#}}
\def\d{  {\rm d}  }

\def\ex{{\rm e}}
\def\lam{\lambda}
\def\hchi{{\raise 0.8mm \hbox{$\chi$}}}
\def\Beven{B_{{\rm e}} }
\def\Bodd{B_{{\rm o}} }
\newcommand{\eq}{\begin{equation}}
\newcommand{\en}{\end{equation}}
\newcommand{\ie}{{\it i.e.}}

\def\sst{\scriptscriptstyle}
\def\sct{\scriptstyle}
\def\ds{\displaystyle}
\newcounter{sublabel}
\outer\def\topic#1{\par{\vskip0pt plus.2\vsize\penalty-250
                        \vskip0pt plus-.2\vsize\bigskip\vskip\parskip
      \onward\setcounter{sublabel}{1} \message{#1}\vspace{0.5cm}
      \leftline{{\large\bf \thesection. #1}}\nobreak\smallskip  \noindent}}
\renewcommand{\thesubsection}{\thesection.\alph{sublabel}}
\def\subonward{\addtocounter{sublabel}{1}}
\outer\def\subtopic#1{\par{\vskip0pt plus.05\vsize\penalty-250
                           \vskip0pt plus-.05\vsize\bigskip\vskip\parskip
   \vspace{0.25cm}
        \leftline{{\it \thesubsection\/ #1}}\subonward\nobreak\noindent}}
\def\sR{{\sst R}}
\def\hep#1#2#3#4{hepth$\#$}

\begin{document}
\setlength{\unitlength}{0.15cm}
\thispagestyle{empty}

\hfill              \begin{tabular}{l} {\bf \hep-th/9406100} \\
                                        {\sf BRX-TH--355} \\
                                        {\sf BOW-PH--102} \\
                                     \end{tabular}

\vspace{1.7cm}

\begin{center}
\begin{tabular}{c}
{\LARGE The String Calculation of Wilson Loops} \\[0.3cm]
{\LARGE   in Two-Dimensional Yang-Mills Theory }
\end{tabular}
\vspace{1.5cm}

\setcounter{footnote}{1}
{\large Stephen G. Naculich,$^*$
 Harold A. Riggs,\footnote{Supported in part by  the DOE under grant
               DE-FG02-92ER40706}
 and Howard J. Schnitzer$^\dagger$}
\vspace{0.5cm}
{ \normalsize \sl
\begin{tabular}{ll}
\kern 0.0em \begin{tabular}{c}
$^{*}$Department of Physics \\
 Bowdoin College  \\
 Brunswick, ME 04011
\end{tabular}    &     \begin{tabular}{c}
                          $^\dagger$Department of Physics \\
                                 Brandeis University  \\
                               Waltham, MA 02254
                                    \end{tabular}   \\[0.5cm]
naculich@polar.bowdoin.edu &
    \begin{tabular}{r}
          hriggs \\
         schnitzer
   \end{tabular}
    \kern -0.95em { @binah.cc.brandeis.edu}
\end{tabular} }\\[1.0cm]

{\normalsize \sf June 1994}

\end{center}

\vfill
\begin{center}
{\sc Abstract}
\end{center}

\begin{quotation}
We demonstrate that the large $N$ expansion of
Wilson loop expectation values
in $\son$ and $\spn$ Yang-Mills theory
on orientable and nonorientable surfaces
has a natural description as a weighted sum
over covers of the given surface.
The sum takes the form of the perturbative expansion
of an open string theory.
The derivation makes contact with the classification
of branched covers by Gabai and Kazez.
Comparison with the analogous results for the chiral sectors of
QCD$_2$ is instructive for both cases.
\end{quotation}
\vfill

\setcounter{page}{0}
\newpage
\setcounter{page}{1}
\setcounter{section}{0}

\topic{Introduction}

The strong form of the recent claims that
QCD\cite{gross,minahan,gt,twist}
and a variety of other gauge theories\cite{nrsstring,ram,dls}
in two dimensions are string theories requires that
all physical observables admit a string representation.
Therefore, the expectation values of Wilson loops,
which form a complete set of gauge-invariant observables
for pure Yang-Mills theory formulated on any surface,
must have a string interpretation.
Early work on ${\rm U}(N)$ gauge theories on the plane\cite{early}
made it apparent that such a representation was possible, and recent work of
Gross and Taylor\cite{twist} gave a detailed, all-orders in $1/N$, string
interpretation of Wilson loops for $\sun$ gauge theory on any
orientable surface.

The purpose of this paper is to
establish the string interpretation of Wilson loops
in $\son$ and $\spn$ gauge theories
on any orientable or nonorientable surface $\cS$.
In particular, we exhibit the expectation value
of any non-intersecting Wilson loop as a weighted sum
over those continuous maps from open worldsheets $\cW$ to $\cS$
that send the boundary of $\cW$ to the Wilson loop in $\cS$;
this key result is expressed in eq.~(\ref{keyresult}).
The sum is organized as the perturbative expansion
of an open string theory.
The derivation makes contact with the relatively recent work
of Gabai and Kazez in which
the classification of branched covers of arbitrary surfaces is
completed.\cite{gk}

This string interpretation
of $\son$ and $\spn$ Wilson loop expectation values
is similar to that already given\cite{twist} for $\sun$ Wilson loops.
For $\son$ and $\spn$, however, the interpretation
allows sheets of $\cW$ to extend over the surface $\cS$
on either side of the Wilson loop
even if they terminate above the Wilson loop (\ref{keyresult}),
while for a single chiral sector of $\sun$,
the sheets ending at the boundary extend only to
one side of the Wilson loop (\ref{chiralresult}).
Thus, the interpretation for $\son$ and $\spn$
is similar to that for the coupled sector of $\sun$
but in a simpler context in which a string representation
for the Wilson loop expectation values
can be written explicitly and compactly.
Comparison of the $\son$ and $\spn$ results with the analogous
result for the string expansion of a single chiral sector
of an $\sun$ Wilson loop
clarifies certain aspects of the latter case,
and illuminates the relation between
the gauge and string pictures for any gauge group.

We do not address intersecting Wilson loops in this paper,
although they are undoubtedly important for a full understanding
of Yang-Mills string theory.
Because $\son$ and $\spn$ tensor products involve contractions
of Young tableau cells,
a string description of $\son$ and $\spn$ intersecting Wilson loops
is comparable in difficulty to intersecting loops
in the {\it coupled sector} of $\sun$ Yang-Mills theory.
A closed form expression for these observables is
simply not possible using the techniques of this paper.

In section {\it 2},
we express certain partition functions and Wilson loop expectation values
in the ``gauge basis,''
\ie, in terms of sums over gauge-group representations.
The classification and counting of maps,
and the weights assigned to these maps by Yang-Mills theory comprises
section {\it 3}.
In section {\it 4},
we reformulate the results of section {\it 2}
in the ``string basis,''
\ie, in terms of sums over symmetric-group conjugacy classes,
which is the natural basis for the string interpretation of these
quantities.
Finally, section {\it 5}\/ illustrates the string-theoretic
calculation of $\son$ and $\spn$ Wilson loops
in several examples.

\topic{Wilson loops in Gauge Theory}

The basic ingredient needed for
the calculation of Wilson loop expectation values
in two-dimensional Yang-Mills theory
is the partition function on an arbitrary surface with boundary.
After recalling the known expressions for this partition function
as a sum over representations
of the relevant gauge group\cite{fine,wit,bt}
(as well as various gluing formulae for partition functions
and Wilson loops on orientable surfaces),
we use it to calculate gluing formulae for partition functions
on nonorientable surfaces and
for Wilson loop expectation values along nonorientable curves.
We also display the explicit formulae needed
to write these expressions as asymptotic expansions in $1/N$.

\subtopic{Partition functions on surfaces with boundaries}

Consider a surface whose boundary consists of $b$ disjoint closed curves.
The value of the Yang-Mills partition function on this surface
depends on the boundary conditions,
which are specified by the holonomies of the gauge field
\eq
U = P \exp\left( i \oint A_\mu \d x^\mu \right)
\label{holonomy}
\en
around each of the boundary curves.
We denote by $  Z(\cS; \setU ) $
the partition function of a surface $\cS$
with holonomies $ \setU $ around the $b$ boundary curves.
The surface may be either orientable or nonorientable.

The partition function $  Z(\cS; \setU ) $
is a class function of each of the group elements $U_j$,
and so may be expanded in terms of the
characters $\gchi_R (U_j)$ of the gauge group,
which form a basis for the class functions,
\eq
Z(\cS; \setU )
=  \sum_\setR  \cZ(\cS; \setR )  \prod_{j=1}^{b} \gchi_{R_j} (U_j) .
\label{characterexpansion}
\en
The sums run over all finite-dimensional, irreducible representations
of $G$.
The coefficients $ \cZ(\cS; \setR ) $
vanish unless the representations $R_j$ associated
with the boundary components all coincide\cite{wit}
\eq
 \cZ(\cS; \setR )
= \cZ(\cS; R_1, \ldots, R_1) \prod_{j=2}^{b} \delta_{R_1,R_j} ,
\label{sumoverreps}
\en
so that
\eq
Z(\cS; \setU )
=  \sum_R  \cZ(\cS; R,\ldots,R )  \prod_{j=1}^{b} \gchi_{R} (U_j) .
\label{onesumoverreps}
\en
The vanishing conditions (\ref{sumoverreps}) are
essential for the string interpretation, as we shall see later.
For an orientable surface, the coefficients
in this expansion are well known to be\cite{mr,fine,wit,bt}
\eq
\cZ(\cS; \sameR)
= (\dim R)^\Eu  \exp\left( -\lam A C_2 (R)\over 2N \right),
\quad \quad {\rm for~}\cS{\rm~orientable}
\label{orientablepartfcn}
\en
where $\Eu = 2 - 2h - b$ is the Euler characteristic
of the surface ($h$ being the number of handles),
$A$ is its area,
$\sqrt{\lam/N} $ is the gauge coupling constant,
and ${\rm dim}R$ and $C_2 (R) $ denote
the dimension and quadratic Casimir of the representation $R$.
For tensor representations of the classical Lie groups,
the quadratic Casimir is given by
\eq
C_2 (R) = fN \left[ r - U(r) + {T(R) \over N} \right],
\en
where
\eq
\begin{array}{rcl}
f &=&\cases{ 1 & for $\sun$ and
               $\son$, \cr {1 \over 2} & for $\spn$, \cr}\\[0.5cm]
U(r) &=& \cases{  r^2/N^2 &  for $\sun$, \cr
	              r/N &  for $\son$, \cr
		    - r/N &  for $\spn$, \cr} \\[0.5cm]
T(R) &=& \ds \sum_{i=1}^{{\rm rank~}G} \, \ell_i (\ell_i + 1 - 2i)
     = \sum^{\col_1}_{i=1} \, \ell^2_i - \sum_{j=1}^{\ell_1} \col^2_j \; ,
\end{array}
\label{things}
\en
with $\ell_i$ ($\col_j$) denoting the row (column) lengths of the Young
tableau corresponding to the representation $R$,
and $r$ the number of cells in the tableau.
All we need to know about the dimension of $R$ is
that ${\rm dim} R = \chi_R^G(\bfid)$.

The unique nonorientable surface with $b$ boundary curves and
Euler characteristic $\Eu=2-q-b$
may be constructed by gluing together $2h+q^{\prime}+b-2$ three-holed spheres
with $h\geq 0$ handles and $q^{\prime}>0$ cross-caps,
provided that $2h+q^{\prime} = q$.
Since a surface with a handle and at least one cross-cap is
homeomorphic to one with three cross-caps, we may remove all handles in
favor of cross-caps or remove all but one or two cross-caps in favor of
handles without changing the topology of the surface.
Any such construction allows one to evaluate the partition function,
yielding\cite{wit}
\eq
\cZ( \cS; \sameR)
= (\sign_R)^{q} (\dim R)^\Eu  \exp\left( -\lam A C_2 (R)\over 2N \right),
\quad\quad  {\rm for~}\cS{\rm~nonorientable}
\label{nonorientablepartfcn}
\en
where $\sign_R =+1$ ($-1$)
if there is a symmetric (anti-symmetric) invariant in $R \otimes R \to \CC$,
and $\sign_R = 0$ if $R \neq \overline{R}$.
This last fact means that,
for nonorientable surfaces, the sum (\ref{onesumoverreps})
is restricted to self-conjugate representations.
All representations of $\spn$
and all tensor representations of $\son$ with $k_1 < N$
are self-conjugate,
and for these representations,
$\sign_R $ is given by $ \sign^r$,
with $\sign = 1$ for $\son$ and $\sign = -1$ for $\spn$,
and with $r$ the number of cells in the Young tableau $R$.
For $\sun$,
the only self-conjugate representations that contribute to
the $1/N$ expansion are {\it composite} representations of the form
$R = \overline{S} S$ (in the notation of ref.~\dcite{gt}).
Since
$\sign_R = (-1)^{r(N-1)}$ for any
$\sun$ representation,\footnote{An extensive discussion of these signs
for all Lie algebras may be found in ref.~\dcite{signref}.}
it follows that $\sign_{\overline{S} S} = 1$.

\subtopic{Partition functions on surfaces without boundaries}

Consider two surfaces $\cS_1$ and $\cS_2$, each with a single boundary
component. The partition function of the closed surface
obtained by gluing these surfaces together along their boundaries
may be found by multiplying their respective partition functions
and integrating over the boundary holonomy\cite{wit}
\eq
\int   \d U   ~ Z(\cS_1; U) Z(\cS_2; U^{-1})
= \sum_{R} \cZ(\cS_1; R) \cZ(\cS_2; R),
\label{gluedpartfcn}
\en
where we have used the character orthonormality relation
\eq
 \int \d U ~ \gchi_{R_1} (U)  \gchi_{R_2} (U^{-1})  =  \delta_{R_1,R_2} .
\label{charorthonormality}
\en

As another example,
consider a single surface $\cS$ with  one or two boundaries.
Each boundary can be glued to itself by identifying antipodal
points. This yields a closed surface with (at least) one or two cross-caps,
respectively. (In fact, any nonorientable surface can be constructed from
an appropriately chosen orientable surface with one or two boundaries by this
means.) With the boundary holonomy given by $U^2$ in the one-boundary
case, the partition function for a closed surface with
a cross-cap insertion is given by
\eq
   \int \d U ~ Z(\cS; U^2) = \sum_R \sign_R \cZ(\cS; R) ,
\label{ccinsert}
\en
where we have used
\eq
  \int \d U ~ \gchi_R(U^2) = \sign_R .
\en
This last equation is a consequence of the identity
\eq
\gchi_R(U^2) = \gchi_{(R\times R)_{\rm sym}}(U)
             - \gchi_{(R\times R)_{\rm anti}}(U) ,
\label{saident}
\en
and eq.~(\ref{charorthonormality}),
since
$\eta_R= {N^+_{RR}}^0 - {N^-_{RR}}^0$,
where
${N^\pm_{RR}}^{R^\prime}$
denotes the tensor product multiplicity of $R^\prime$ in
the symmetric ($+$) or anti-symmetric ($-$) part of $R \otimes R$.

Given a surface with two boundaries, the boundaries can be glued to
themselves or to each other by antipodes, but with the same result
(the partition function of a closed surface with a twisted handle
or Klein bottle insertion)
\eq
\left.
\begin{array}{r} \ds \int \d U_1 \d U_2 ~ Z(\cS; U_1^2, U_2^2) \\
  \ds \int \d U_1 \d U_2 ~ Z(\cS; U_1 U_2, U_1 U_2) \end{array}
\right\}
= \sum_R \delta_{R,\overline{R}} \cZ(\cS; R,R) .
\en
Comparison with the regular-handle insertion formula,
\eq
   \int \d U ~ Z(\cS; U, U^{-1}) = \sum_R \cZ(\cS; R,R),
\en
illustrates the implicit convention that $Z(\cS; \setU)$ is written
with a uniform choice of relative orientations on the boundaries.

\subtopic{Wilson loops}

A Wilson loop is the trace, taken in a given representation $R$,
of the gauge-field holonomy (\ref{holonomy}) for a given closed curve,
\ie, the character $\gchi_R (U)$ of the gauge-group element $U$.
The expectation value of a product of Wilson loops on a surface
may be calculated by cutting the surface along the curves of each loop,
sandwiching $\gchi_R(U)$ between the partition functions
of the resulting surfaces with boundaries, and then integrating over the
holonomies around the boundary curves.\cite{wit,bt}

Consider first a single homologically trivial Wilson loop,
\ie, one which cuts a surface $\cS$
into two disjoint surfaces $\cS_1$ and $\cS_2$.
The expectation value of the Wilson loop
associated with representation $R$ is given by\cite{bt}
\begin{eqnarray}
W_R
& = & \int \d U ~ Z(\cS_1; U) \,  \gchi_R(U) \,  Z(\cS_2; U^{-1}) \nonumber\\
& = & \sum_{R_1,R_2} \NG{R_1}{R}{R_2} \cZ(\cS_1; R_1) \cZ(\cS_2; R_2)  ,
\label{homologtrivial}
\end{eqnarray}
where
$ \NG{R}{S}{T}$ are tensor product multiplicities. They satisfiy
\eq
\gchi_R (U) \gchi_S(U) = \sum_T  \NG{R}{S}{T}  \gchi_T (U).
\label{charproduct}
\en
For $\cZ(\cS_i; R_i)$ in eq.~(\ref{homologtrivial}),
one should use either (\ref{orientablepartfcn}) or
(\ref{nonorientablepartfcn}),
according to the orientability of $\cS_i$.

If the Wilson loop is homologically non-trivial and
orientable,\footnote{Consider a narrow band along the closed curve
of the Wilson loop: if that band is a cylinder the curve is termed
orientable, while if it is a M\"obius band the curve is termed nonorientable.}
cutting the surface along the loop results in a single surface
with {\em two} boundary components,
whose partition function is $Z(\cS; U_1, U_2)$.
The expectation value of the Wilson loop, calculated as just described,
is\cite{bt}
\eq
\int \d U ~ Z(\cS; U, U^{-1})  \,  \gchi_R(U)
=  \sum_{R^\prime} \NG{R^\prime}{R}{R^\prime}
 \cZ(\cS; R^\prime, R^\prime). \label{hnontriv}
\en
Since the tensor product algebras of the classical groups
satisfy certain conservation laws
(for $\sun$ the number of Young tableau cells is conserved mod $N$,
while for $\spn$ and $\son$ the number of cells is conserved mod $2$\footnote{
While this is not completely true for $\son$ with $N$ odd,
the only exceptions occur for tableaux with order $N/2$ cells,
which do not appear in asymptotic $1/N$ expansions.}),
a single homologically-nontrivial (orientable) Wilson loop for a
representation $R$ is only nonvanishing for representations whose
tableaux have zero mod $2$ (for $\son$ and $\spn$)
or zero mod $N$ (for chiral $\sun$) cells.

On a nonorientable surface,
Wilson loops on nonorientable curves are
also possible, and cutting along such a curve results in
a single surface with {\em one} boundary component.
The paradigmatic case is the Wilson loop on the projective plane
(pictured as a disk with antipodal boundary points identified)
that connects antipodes.
Using (\ref{saident}),
the expectation value of such a Wilson loop is
\eq
\int \d U ~ Z(\cS; U^2) \gchi_R(U)
=  \sum_{R^\prime}  ({N^+_{R^\prime R^\prime}}^R - {N^-_{R^\prime R^\prime}}^R)
    \cZ(\cS; R^\prime),
\label{weird}
\en
which reduces correctly to (\ref{ccinsert}) for $R = 0$.
Note that (\ref{weird}) implies that such Wilson loops vanish for all gauge
groups unless the number of cells in the tableau $R$ is even, and that
it is nonvanishing for all tableaux $R$ with an even number of cells,
even in a chiral sector of the $\sun$ string expansion.
The sum over $R^\prime$ provides a finite number of terms for
each chiral $\sun$ sector
but an infinite number of terms for the $\son$ and $\spn$ expectation values.

Finally, it is interesting to compare the orientable Wilson
loop~(\ref{hnontriv}) with the analogous nonorientable Wilson loop
around a twisted handle
\eq
 \int \d U_1 \d U_2 ~ Z(\cS; U_1 U_2, U_1 U_2) \gchi_R(U_1 U_2)
  =  \sum_{R^{\prime}} \NG{R^\prime}{R}{\overline{R^\prime}}
                      \cZ(\cS; R^\prime, R^\prime).
\label{twistedhandle}
\en
If $\cS$ is orientable, these results are identical
for $\son$ and $\spn$.
Both (\ref{hnontriv}) and (\ref{twistedhandle})
vanish in each chiral $\sun$ sector.

The same procedures apply to an arbitrary
number of non-intersecting Wilson loops.
Although explicit and well-suited for computations,
the Wilson loop formulae in this section are ill-suited for a
direct string interpretation.  As we show in section {\it 4}
a different basis for the partition functions on surfaces with
boundary is needed to make the string theoretic
character of two-dimensional Yang-Mills theory evident.
The group-theoretic selection rules mentioned
above have natural topological implementations,
which we also illustrate in section {\it 4}.

\topic{How Yang-Mills String Counts Covers}

We will show in the next section that the terms in the large $N$ expansion
of $\son$ and $\spn$ Wilson loop expectation values on a surface $\cS$
correspond to certain classes of covers of $\cS$.
In this section we describe these covers
and the weights that will be given to them via this correspondence.

While the natural boundary condition for a gauge theory
on a surface $\cS$ with boundary
is specified by the gauge-field holonomies around the boundary curves,
the natural boundary condition in a string theory on such a surface
can only involve the different ways
that the boundary of a string worldsheet $\cW$ can be mapped to $\cS$.
We require the boundary of the covering surface $\cW$ to be mapped to
the boundary of $\cS$: since only the total area $A$ of $\cS$ is
preserved under area-preserving diffeomorphisms, and since maps
are to be weighted by the area of the cover, all maps that send
part of the boundary of $\cW$ to the interior of $\cS$ or that
have finite-length folds will not
have invariant weights under this symmetry and so are disallowed.
This requirement implies that every boundary must be covered by exactly
the same number of sheets and that the area of $\cW$ must be an integral
multiple of $A$.

The possible $r$-fold covers of a boundary component of $\cS$ by
boundary components of arbitrary covering surfaces are in 1--1
correspondence with the partitions of $r=\sum_i l_i$ into integers $l_i$:
each element $l_i$ of a given partition of $r$ is associated with
the unique, connected $l_i$-fold cover of a circle by a circle.
Given a labeling of the sheets of the cover,
the lifts to $\cW$ of circuits of each boundary of $\cS$
yield a permutation of the labels for each boundary, with the integers
$l_i$ giving the lengths of the cycles in the permutation.
Since any labeling will do,
we see that the possible string boundary conditions
for an $r$-fold cover correspond precisely to the conjugacy classes of
the permutation group of $r$ symbols, $S_r$.

We now consider the set of $r$-sheeted covers of a surface $\cS$
with specified conjugacy classes $\kap_j$ ($j=1,\ldots, b$) indicating the
boundary covering on each boundary component of $\cS$.
All such covers that map boundary to boundary factorize into a composition
of a true cover, a ramification-point identification map,
and a pinch map,\cite{factorize} as we shall now illustrate.
The following is a slight generalization of the technique
deployed by Gabai and Kazez\cite{gk} in their classification of branched
covers to include the case of surfaces with boundary.

One can construct any foldless, boundary-to-boundary continuous cover of $\cS$
from $r$ copies $\cS_i$ ($i=1,\ldots, r$) of $\cS$
as follows.
Initially consider the completely disconnected
cover given by the $r$ identity maps $\iota: \cS_i \rightarrow \cS$.
It is useful to picture each of the $r$ copies $\cS_i$ hovering above
(and surrounding, as Matrushka dolls do one another)
$\cS$ such that all handles
(\ie, all homology basis cycles) of $\cS_i$ are coaxial with
those of $\cS$.
In this way we can view
the label $i$ as the height of $\cS_i$ above $\cS$ (at height zero) in the
space $ [0,r] \times \cS$
where $[0,r]$ is a real line segment (heuristically of length $r$).
The boundary condition on each boundary component is just that
associated with the identity element (and conjugacy class) of $S_r$.
Any true cover\footnote{A cover is true if the inverse image of every point of
$\cS$ contains exactly $r$ points.} with this trivial boundary condition may
then be obtained from the disconnected cover just described
by first selecting an
{\em appropriate} set of homologically nontrivial curves in $\cS$
(they are necessarily closed loops and may need to wind several times
around homology cycles),
by then selecting an {\em appropriate} pair of sheets
above a chosen initial point on each curve and cutting along the lifts of
each curve that begin on the two chosen sheets above the initial
point,
and finally by reconnecting the upper sheet on one side with
the lower on the other, and vice versa.
(If two curves in $\cS$ intersect, then the lifts
may change their height during a circuit of the curve so that the
designation upper and lower may change as one moves along one of the lifts.)
One cannot, of course, carry out this construction for all choices of
curves and pairs of sheets: thus arises the nontrivial counting problem
for true covers. (The problem of counting
covers with trivial boundary conditions is precisely equivalent to the
celebrated problem\cite{sftext} of counting the true covers of closed
surfaces.)
Note that the curves chosen in $\cS$ are, by construction, the projections
of the self-intersections of the {\em immersion} of the cover in the
space $[0,r]\times \cS$ (\ie, they do not correspond to actual intersections
of $\cW$). Following ref.~\dcite{gk}, we will refer to the projections
to $\cS$ of the self-intersection lines of the immersion of any cover
(in this space) as {\em double arcs}.

Arbitrary true covers with nontrivial boundary conditions may be obtained
from the true covers with trivial boundary conditions just constructed by
further surgery above curve segments in $\cS$ that begin
and end on boundaries.  Above {\em one} end of each such curve choose
two sheets, cut along the two lifts, and reconnect upper on one side to
lower on the other. In order for a curve to begin and end on the {\em same}
component and produce a nontrivial effect via such surgery, the curve must
cross (at least) one of the double arc projections used to construct the
initial true cover. A circuit of a given boundary component will now induce a
permutation of the labels of the cover given by a product of transpositions,
one for each line segment that begins or ends on the given boundary
component.

An arbitrary simply-branched cover\footnote{A cover is branched
if the inverse image of all but a finite number of points
of $\cS$ contain exactly $r$ points with the inverse
image of the remaining points containing from $1$ to $r-1$ points;
a cover is {\em simply} branched if all the points whose inverse images
contain fewer than $r$ points contain exactly $r-1$ points.}
may now be constructed by surgery on any of the true covers just
constructed by first selecting an arbitrary set of open line segments in $\cS$,
by then selecting an arbitrary  pair of sheets
above {\em one} end of each line segment
and cutting along the two lifts of each line segment
that start above the selected end,
and finally by reconnecting the upper sheet
on one side above each point along the segment in $\cS$ with the
lower sheet on the other side, and vice versa.
This is only possible
if the end points of the two lifts are identified in $\cW$,
not just the immersion. In this way the
ends of the line segment become branch points.
This construction makes it
apparent that not only is the number of simple branch points on a
closed surface even, but that simple branch points necessarily come
in pairs connected by double arc projections.
(This is true for any {\em fixed} configuration of double arcs:
by bringing pairs of double arcs together one can disconnect a given
pair of branch points, but in the process they will necessarily become
reconnected, again in pairs, to other branch points.)
An order $s$ multiple branch point may similarly be constructed by choosing
$s-1$ line segments all starting at a single point, and performing the
reconnection surgery, as described above, along each line segment.
In addition, one may choose line segments starting on a boundary and ending
at any interior point, choosing a pair of sheets above the boundary endpoint
and performing the reconnection surgery as above.
One sees that, in all cases, the line segments on $\cS$ are again the
projections of the self-intersections of the immersion of the cover
in the space $[0,r]\times \cS$.  If
$\cS$ is orientable, the double arcs that connect branch points
correspond to a specification of analytic branch cuts.
The above topological picture makes it clear, however,
that double arc projections form
a useful and well-defined extension of the idea of branch cuts to
nonorientable surfaces.

The claim of Gabai and Kazez\cite{gk}
is that any branched cover can be represented by such
a branched immersion of the cover in $[0,r]\times \cS$. However,
many of these constructions are topologically equivalent. Gabai and
Kazez show that one can analyze the topological equivalence  of
different covers entirely via shifts of the double arc projections on $\cS$,
and find that any branched cover must reduce to one of a number of
standard forms. They prove that the branched covers of any surface
(orientable and nonorientable) with a given number of branch points
are classified by the degree of the  cover $\nu$ (essentially,
the number of sheets) and by the homomorphism
$\nu^* : \pi_1(\cW) \rightarrow \pi_1(\cS)$ of fundamental groups
induced by the cover $\nu$. This means that the topological equivalence
classes of branched covers of a surface with a given number of
branch points $\{p_i\}$ are
classified by the conjugacy classes of subgroups of $\pi_1(\cS)$ and
not $\pi_1(\cS-\{p_i\})$.

Branched covers do not exhaust the types of
foldless, boundary-to-boundary continuous covers.
The map $\mu: \cW \rightarrow \cW^\prime$ that shrinks any open set
to a point can clearly be composed with any of the branched covers
$\nu : \cW^\prime \rightarrow \cS$ to
form a continuous cover in which the inverse image of certain
{\it pinch points} in $\cS$ are open sets in $\cW$. The open
set may contain entire handles or cross-caps. In order for the
dynamical weight of a map to be given by the only invariant under
area-preserving diffeomorphisms, the total area of $\cS$,
the area of the open set must be
set to zero: we only allow infinitesimal pinches.

For a given set of conjugacy classes $\kap_j$ ($j=1,\ldots, b$)
describing the covering of the $b$ boundary components of $\cS$,
all the possible covers may in principle be constructed as described above.
The sum that we will associate with the covers that
satisfy this boundary condition runs over all true covers as well as all
possible covers with an arbitrary number and configuration of simple
branch points and an arbitrary number and configuration of insertions
of pinched handles (for $\sun$) or pinched cross-caps (for $\son$ and $\spn$).

If the Euler characteristic of $\cS$ is not zero,
then each of the $r$-sheeted covers just described appears as the leading
member in a {\em twist} family of covers that must also be included.
The other members of each family of
covers contain the leading cover with additional features inserted
as prescribed by the form of the twist
operator.\cite{twist,ram,nrstwist}
This element of the Frobenius algebra of $S_r$,
to be defined more precisely in the next section,
is a sum over all elements of $S_r$,
with each element corresponding to a multiple branch point of the cover.
On a surface with Euler characteristic $\Eu$,
exactly $|\Eu|$ twist operators (or their inverses) will appear,
so that we must consider covers with the
insertion of $|\Eu|$ (or fewer) fixed but arbitrary simple or multiple
branch points.
While the sum over members of the twist family is topologically
obscure (but see ref.~\dcite{cmr}),
the sum over families (\ie, over the leading member of each family) runs
over a complete, natural, and unrestricted topological class:
all continuous, simply-branched
covers composed with arbitrary pinch maps of tubes and
handles (for $\sun$) or cross-caps (for $\son$ and $\spn$).

For true covers the idea of counting the topologically distinct,
connected, $r$-sheeted covers of $\cS$ by counting homomorphisms
$h_\nu: \pi_1(\cS) \rightarrow S_r$ from the fundamental group to
the symmetric group induced by the cover (and the proof
that this is equivalent to counting conjugacy classes of subgroups
of $\pi_1(\cS)$) goes back to Reidemeister\cite{rmeister} and is given a
detailed exposition by Seifert and Threlfall in section 58 of
their celebrated textbook.\cite{sftext} There it is shown that the
{\em connected}, $r$-sheeted, true covers of a surface are in one-to-one
correspondence with the distinct-under-relabeling, {\em transitive}
solutions of the generating relation of $\pi_1$ for the surface given by
elements of $S_r$. They have shown, therefore, that the number of
connected $r$-sheeted covers is given by
\eq
\sum^{*}_{
 	{{ {a_1, \ldots, b_h                \in S_r} \atop
       {b_1, \ldots, b_h                \in S_r} } \atop
	{c_1, \ldots,  c_{q^\prime}    \in S_r} } }
\delta ([a_1,b_1]\ldots [a_h,b_h] c^2_1 \cdots  c^2_{q^{\prime}}),
\label{sfres}
\en
in which $[a,b]=aba^{-1}b^{-1}$ is the the group commutator and
in which the symmetric group delta function $\delta (\rho)$ is defined by
\eq
\delta  (\rho )
= \left\{
\begin{array}{lll}
1 & {\rm if} & \rho ~ = ~{\rm identity},  \\
0 & {\rm if} & \rho ~ \neq ~{\rm identity} .
\end{array}
\right.
\en
The star in (\ref{sfres}) indicates that we only include transitive
solutions and that we must avoid overcounting due to
equivalence under sheet relabeling, and $h$ and $q^\prime$ are the number
of handles and cross-caps respectively, as described in the previous section.
By including all (not just the transitive) homomorphisms,
one obtains a sum that counts all (not just the connected) covers.
If, in addition,
one does not take account of the equivalence under relabeling,
the overcounting factor\cite{gt} is
$r!/|S_\nu|$, with $|S_\nu|$
denoting the number of homeomorphisms $\rho$ of $\cW$ for which
$(\nu \circ \rho )(p) = \nu (p)$ for all points $p\in \cW$.
Therefore,
the sum over all topologically distinct, possibly disconnected true covers
$\Cover(\cS)$ of the closed surface $\cS$,
with each cover given the weight $1/|S_\nu|$,
is given by
\eq
\sum_{ \nu\in\Cover(\cS) } \; {1 \over |S_\nu|}
= \sum_{
 	{{ {a_1, \ldots, b_h                \in S_r} \atop
       {b_1, \ldots, b_h                \in S_r} } \atop
	{c_1, \ldots,  c_{q^\prime}    \in S_r} } } \;
{ 1 \over r!} \,
\delta ([a_1,b_1]\ldots [a_h,b_h] c^2_1 \cdots  c^2_{q^{\prime}} ) .
\en

In Yang-Mills string theory, however,
we need to calculate a weighted sum over a wider class of covers.
Consider a surface $\cS$ with
$b$ boundary components,
$h$ handles, and
$q^\prime \ge 0$ cross-caps
(so that $\Eu = 2-2h-q^\prime-b$).
Let $\Covers$ denote the set of all
possibly-disconnected, foldless, boundary-to-boundary
$r$-fold continuous, true and simply-branched covers of $\cS$,
with the boundary mapping given by the conjugacy classes
$\setkap$ along the boundary curves,
with $\nbr$ simple branch points at {\em fixed} locations on $\cS$,
with $k$ pinched handles (for $\sun$) or cross-caps (for $\son$ and $\spn$),
and with fixed multiple branch points
whose ramification points are chosen so that their images
under the homomorphism
$h_\nu$ are given by the elements
$\rho_l \in S_r$ ($l=1,\ldots, \Eu$).
The claim made in refs.~\dcite{gt,twist,nrsstring}
is that the sum over such covers,
each weighted by the topological symmetry factor $1/|S_\nu|$,
and by a dynamical weight $f_\nu(\cA, N)$ to be specified below,
is given by
\eq
\begin{array}{l}
\ds \sum_{ \nu\in\Covers }  \; {f_\nu(\cA, N)\over |S_\nu|}
   \hfill  \\
\ds \hfill ~\quad
         = \sum_{ { { {a_1, \ldots, a_h \in S_r} \atop
       {b_1, \ldots, b_h                \in S_r} } \atop
       {c_1, \ldots,  c_{q^\prime}    \in S_r} } \atop
       {p_1, \ldots,  p_\nbr            \in P_r}  } \;
\sum_{s_j  \in {\kap_j} } \;
{ f_\nu(\cA, N) \over r!} \,
\delta ( [a_1,b_1]\ldots [a_h,b_h] c^2_1 \cdots  c^2_{q^{\prime}}
 p_1 \cdots p_\nbr s_1 \cdots s_b \prod_{l=1}^{\Eu}\rho_l ),
\end{array}
\label{weightsum}
\en
where $P_r$ is the conjugacy class of permutations in $S_r$ that
interchange only two elements.

In light of the classification theorem for branched
covers\cite{gk} the presence of the
generating relation of $\pi_1(\cS-\{ p_i \})$ deserves comment. The
transposition assigned to a homotopy curve that encircles
a single simple branch point (and nothing else) by the homomorphism
$h^\prime_\nu : \pi_1(\cS-\{p_i\}) \rightarrow S_r$
depends on the choice of
a base point for the homotopy curve relative to the locations of the
double arc projections.
The base point can always be chosen so that
the the numbers interchanged by the transposition
assigned to this homotopy curve also label the sheets
connected by the ramification point in $\cW$ above the branch point.
In this case, the sum over each $p_i\in P_r$ in
eq.~\ref{weightsum} becomes a sum over the pairs of sheets connected by the
ramification point above a given branch point. Terms can arise in this
sum which correspond to the insertion of a fixed number of ramification
points into
a {\em connected} component of the underlying true cover in different
ways. The theorem of ref.~\dcite{gk} implies that all such covers
are topologically equivalent. The proof appeals to certain lemmas
which show that one can move the simple
branch points across double arcs so that one can remove all the intersections
of each double arc (connecting branch points) with all other double arcs.
The homotopy curve around each pair of branch points connected by
a double arc that does not intersect any other double arcs
will be assigned the trivial permutation. In this way one can remove
the product $p_1 \ldots p_i$ from eq.~\ref{weightsum} at the cost of
an integer multiplicity. (This is a reflection of the fact that
area-preserving diffeomorphism invariance of the
gauge theory implies that such covers should appear in the string expansion
with equal weight.)
Since the lemmas of ref.~\dcite{gk} only depend on the local structure
of the cover, the same statement is true
for the the insertion of ramification points that connect components of
the cover unconnected in the underlying true cover. (The difference is
that in the first, connected component, case, the ramification points can
all be moved so that they each connect any given pair of sheets, while
in the second case, this is not possible.)
Therefore, sum~(\ref{weightsum}) is a combined sum over topological classes
and over (possibly topologically equivalent) ramification point
configurations. When combined with the motion of the ramification points
over the surface without changing the sheet connections (which is accounted
for in the weight $f_\nu$) we obtain a sum over all possible locations
and configurations of the ramification points as well
as a sum over topological classes of simply branched covers.

We will now evaluate the weighted sum over covers (\ref{weightsum}).
Let $D_R(\rho )$ be the matrix associated with $\rho \in S_r$ in the
representation of $S_r$ specified by the Young tableau $R$;
let $I_R$ be the identity matrix in this representation.
The symmetric group character of $\rho$
is $\hchi_R (\rho ) = \Tr \, D_R (\rho )$,
and $d_R = \Tr ~I_R $ is the dimension of $R$.
Using Schur's lemma and the fact that all the representations
of $S_r$ are real, it follows that
\eq
    \sum_{a,b \in S_r} \, D_R([a,b]) = \left({r! \over d_R}\right)^2 \: I_R
\en
and that
\eq
\sum_{c \in S_r} \, D_R(c^2) =  {r! \over d_R} \: I_R.
\label{schurident}
\en
The sum over the elements of a conjugacy class yields
\eq
\sum_{s\in \kap} \; D_R (s)
= { \left| \kap \right| \over d_R }  \hchi_R (\kap)  I_R,
\en
a special case of which is\cite{gt}
\eq
\sum_{p\in P_r} \; D_R(p)
= {T(R) \over 2}  I_R,
\label{specialcase}
\en
where $T(R)$ is given by eq.~(\ref{things}).
Writing the delta function as
\eq
\delta  (\rho ) =
{1 \over r!} \: \sum_R \: d_R \: \hchi_R (\rho )
\en
and using eqs.~(\ref{schurident}-\ref{specialcase}),
we may rewrite (\ref{weightsum}) as
\eq
\begin{array}{l}
\ds \sum_{ \nu\in\Covers }  \; {f_\nu(\cA, N)\over |S_\nu|}
   \hfill  \\
\ds \hfill ~\qquad
= \sum_{R\in Y_r} \, {f_\nu(\cA, N) \over r!} \,
\hchi_R (\prod_{l=1}^{\Eu} \rho_l )
\left( r! \over d_R \right)^{q-1}
\left( T(R) \over 2 \right)^\nbr
\prod_{j=1}^{b}  { \hchi_R (\kap_j) \over d_R } |\kap_j|
\end{array}
\label{coversum}
\en
where $q=2h+q^{\prime}$.
The surface is orientable for $q^{\prime} = 0$
and nonorientable for $q^{\prime} > 0$.
(Note that the forms of the sums are identical for orientable and
nonorientable surfaces with the same Euler characteristic.)

The dynamical weight assigned to each cover in Yang-Mills string theory is
\eq
  f_\nu(\cA, N) =  \ex^{-\cA r/2} \; {\cA^{\nbr} \over \nbr!}
(-1)^\nbr  N^{r\Eu - \nbr} \sign^{q^\prime r} \times
\cases{
{1 \over k!} \left(r^2 \cA \over 2 N^2 \right)^k  & for $\sun$ \cr
{1 \over k!} \left( \sign r  \cA \over 2 N   \right)^k  &  for $\son$ and
$\spn$
}
\en
with $\cA = f \lam A$ denoting the dimensionless area of $\cS$.
The exponential factor $ \ex^{-\cA r/2} $ weights each
cover $\cW$ by its area $r \cA$,
as expected from a leading Nambu-Goto term in the action.
For a given configuration of ramification point sheet identifications
(which is being implicitly summed over),
we weight the different locations of the ramification points by
a uniform area measure,
which contributes a factor of $\cA$
for each of the $\nbr$ simple ramification points,
and an overcounting correction factor of $1/\nbr !$ due to
the indistinguishability of the ramification points.
For $\son$ and $\spn$ we allow the covers to contain an arbitrary
number $\ncc$ of infinitesimal cross-caps\cite{nrsstring} mapped to
points, each of which decreases the Euler characteristic of
the cover by $1$; each cross-cap carries a factor of $r \cA$ to
account for its possible location on any of the $r$ sheets of the cover,
with the overcounting  correction of $1/ \ncc!$ to
account for their indistinguishability.
In a chiral $\sun$ sector, an arbitrary number $k$ of infinitesimal
handles and tubes must be allowed,\cite{minahan,gt,twist}
each of which decreases the Euler characteristic of the cover by $2$
and contributes a factor of $ r^2 \cA$ for the location of its ends,
and a factor of $1/k!$ to account for their indistinguishability.
Thus,
for every topologically distinct configuration,
we also count as distinct the different ways
of moving the singular points
(whether ramification points, pinched cross-caps, or pinched handles)
around, but weight them by the area of the cover traversed.
Although a surface with a handle and cross-cap insertion is
homeomorphic with the same surface with three cross-cap insertions, the
area dependence of these two insertion types is different. While no
ambiguity arises since the prescription is
cross-cap pinches for $\son$ and $\spn$ and handle pinches for
$\sun$, this fact seems to require that the two constructions are
not connected by any area-preserving diffeomorphisms.
The weight also includes a factor of $N$
raised to the Euler characteristic of the {\it cover}
($ r\Eu - \nbr - k$ for $\son$ and $\spn$ and
$ r\Eu - \nbr - 2k$ for $\sun$),
which corresponds to the coupling constant of the string genus expansion.
Finally, we assign a factor $-1$ for each simple branch point,
$1/2$ for each {\em pinched} cross cap or handle,
and a factor $\sign$ for each cross-cap (whether pinched or not) in the cover.
(Set $\sign = 1$ for $\son$ and $\sun$, but set $\eta = -1$ for $\spn$).

\topic{Wilson loops in String Theory}

In this section we demonstrate that Wilson loop expectation values
in two-dimensional $\son$ and $\spn$ Yang-Mills theory
have a natural geometric interpretation in terms of the
sum over covering maps just described.
To do so, we introduce a new basis of class functions and
show that the expansion coefficients of
$\son$ and $\spn$ partition functions on surfaces with boundary
in this new basis have a natural string interpretation.

\subtopic{Partition functions on surfaces with boundaries}

In light of the discussion of boundary conditions
found at the beginning of the previous section,
we choose a basis of class functions  $\Upkap (U)$
labeled by conjugacy classes $\kappa$ of the symmetric group $S_r$.
This is possible since the number of conjugacy classes of $S_r$
equals the number of Young tableaux with $r$ cells.
We denote the set of classes of $S_r$ by $\cK_r$.
In particular,
consider the class functions given by the
symmetric-group transforms of the gauge-group characters
\eq
\Upkap (U) = \sum_{\RinY}  \hchi_R (\kap)  \gchi_R (U), \qquad \kinK,
\label{schurdef}
\en
where $\gchi_R (U)$ is the character of $U$ in the representation of $G$
labeled by a Young tableau $R$,
and $\hchi_R(\kap)$ is the character of (any element of the conjugacy
class) $\kap$
in the representation of the symmetric group $S_r$
labeled by the same Young tableau $R$.
$Y_r$ is the set of Young tableaux with $r$ cells.
These class functions form a well-defined basis due to the
symmetric group orthogonality relations:
\begin{eqnarray}
\sum_{R\in Y_r}  \hchi_R (\kap_1) \hchi_R (\kap_2)
         & =  & C_{\kap_1} \delta_{\kap_1, \kap_2} ,  \\
\sum_{\kinK} {1\over C_\kap } \hchi_{R} (\kap) \hchi_{R^\prime} (\kap)
& = &  \delta_{R, R^\prime} .
\label{symmetricorthonormality}
\end{eqnarray}
The integer
\eq
         C_\kap = {r!\over |\kap |} ,
\en
where $|\kap|$ is the number of elements in the conjugacy class
$\kap$, plays a crucial role in the
string sewing formulae that follow.

The partition function of a surface $\cS$
with holonomies $U_j$ around its $b$ boundary curves has the expansion
\eq
Z(\cS;  \setU ) = \sum_{r_i=0}
\sum_{{\kap_i \in \cK_{r_i} \atop {\sct \kern -0.9cm {i=1,\ldots,b}}}}
   \cZ(\cS; \setkap )  \prod_{j=1}^{b} \Up_{\kap_j} (U_j) .
\label{schurexpansion}
\en
Using eq.~(\ref{onesumoverreps}) and
the orthogonality relations
\eq
\int \d U~\Up_{\kap_1} (U) \Up_{\kap_2} (U^{-1})
= C_{\kap_1} \delta_{r_1, r_2} \delta_{\kap_1, \kap_2},
\label{schurorthonormality}
\en
where $\kappa_i \in \cK_{r_i}$,
one finds the relation between the gauge-basis and
string-basis expansion coefficients (letting $r=r_i$ for any $i$)
\begin{eqnarray}
\cZ(\cS; \setkap ) &  = &  \sum_{  \RinY }  \cZ (\cS; \sameR)
\prod_{j=1}^b \delta_{r_i, r} \hchi_{R}(\kap_j) C_{\kap_j}^{-1} ,
\label{schurcoefficients}
\end{eqnarray}
so that $\kap_j \in \cK_{r}$ for all $j$.
The fact that
$\cZ(\cS; \setkap)$ vanishes unless all the $\kap_j$ belong to
the same symmetric group $S_r$ follows from (2.3).
The string meaning of these vanishing conditions is that
the only allowed covers are those that map boundary to boundary, and
the only such maps are those with $r$-sheeted covers over each boundary
component. Other maps will not be invariant under area-preserving
diffeomorphisms.

The coefficients $ \cZ(\cS; \setkap ) $  of the expansion
(\ref{schurexpansion})
for $\son$ and $\spn$ Yang-Mills theory on any surface $\cS$
have a natural interpretation as a sum over covers of $\cS$.
(In ref.~\dcite{twist}, this was demonstrated for the coefficients
of the $\sun$ partition function on an orientable surface.
Note that the quantity $Z(\cS; \sigma_1,\ldots,\sigma_b)$
defined in ref.~\dcite{twist}
differs from $\cZ(\cS; \setkap)$ by the factors $C_{\kap_j}^{-1}$;
this is because we have expanded $Z(\cS; \setU)$
as a sum over conjugacy classes rather than permutations.)
To show this, we first introduce the notion of a twist element $\Omr$,
following refs.~\dcite{twist,ram,nrstwist}.
Due to the orthogonality of symmetric group characters
(\ref{symmetricorthonormality}),
the inverse of  (\ref{schurdef}) is
\eq
\gchi_R (U) =
\sum_{\kinK}  C_\kap^{-1} \hchi_R (\kap) \Upkap (U)\; , \qquad \RinY \; .
\en
The dimension of the $G$-representation $R$  is then
\eq
\dim R
= \gchi_R (\bfid)
=   \sum_{\kinK} C_{\kap}^{-1} \hchi_R (\kap) \Upkap (\bfid)
= {N^r\over r!}  \hchi_{\sR} ( \Omr ),
\en
where $\Omr$ is the $S_r$ group algebra element
\eq
\Omr = {1\over N^r} \sum_{\rho\in S_r} {\Up_{\kap_\rho}} (\bfid)
\; \rho,
\label{omegadef}
\en
different for each group $G$.
Explicit expressions for $\Omr$
are given in ref.~\dcite{twist} for $\sun$
and in ref.~\dcite{ram} for $\son$ and $\spn$.
Since $\Upkap (U) $ is a nonvanishing class function of $S_r$,
$\Omr$ commutes with all the elements of $S_r$,
and is therefore by Schur's lemma
a (non-zero) multiple of the identity matrix in any representation $R$.
Consequently,
\eq
\hchi_{\sR} (\Omr^m)
= (d_{\sR})^{1-m} \left[ \hchi_{\sR} (\Omr) \right]^m
= d_R^{1-m} (r! N^{-r} \dim R)^m,
\label{omega}
\en
where $d_{\sR}$ is the dimension of the symmetric group representation $R$.
We will refer to $\Omr$ or any power thereof as a twist operator.

The coefficients $\cZ(\cS; \setkap)$ (with $\kap_j \in {\cal K}_r$)
for $\son$ and $\spn$,
or for a single chiral sector of
$\sun$,\footnote{ Since $\sign_{\overline{S} S} = 1$ for $\sun$,
we assign $\sign_R =1$ to all representations in a single chiral sector.}
on an orientable or nonorientable surface $\cS$
with $b$ boundary components, $h$ handles, and $q^\prime \ge 0$ cross-caps
(with $q= 2h+q^\prime$ and $\Eu = 2-q-b$)
are given by
(using eqs.~(\ref{schurcoefficients}),
(\ref{orientablepartfcn}), and (\ref{nonorientablepartfcn}))
\begin{eqnarray}
\cZ(\cS; \setkap)
& = &
\sum_{\RinY}
(\sign_R)^{q^\prime}
(\dim R)^\Eu \;
\ex^{-\lam A C_2 (R) /2N}
\prod_{j=1}^{b}  \hchi_R (\kap_j) C_{\kap_j}^{-1}
\nonumber\\
& = &
\sum_{\RinY}
\sign^{{q^\prime}r}
(\dim R)^\Eu \;
\ex^{-\cA r/2} \;
\ex^{ \cA U(r)/2 } \;
\ex^{ - T(R) \cA / 2N  }
\prod_{j=1}^{b}  {\hchi_R (\kap_j) \over r! } |\kap_j|
\nonumber\\
& = &
\sum_{\nbr=0}^{\infty}
\sum_{k=0}^{\infty}
\sum_{R\in Y_r} \, {f_\nu(\cA, N) \over r!}
\hchi_R ( \Omega_r^{\Eu} )
\left( r! \over d_R \right)^{q-1}
\left( T(R) \over 2 \right)^\nbr
\prod_{j=1}^{b}  { \hchi_R (\kap_j) \over d_R } |\kap_j|
\nonumber\\
& = & \sum_{\rho_1, \ldots, \rho_\Eu \in S_r}
\left[ \prod_{l=1}^{\Eu} {1\over N^r}{\Up_{\kap_{\rho_l}}} (\bfid) \right]
\sum_{\nbr=0}^{\infty}
\sum_{k=0}^\infty
\sum_{\nu\in\Covers} \; {f_\nu(\cA, N)  \over |S_\nu|} .
\nonumber\\
\end{eqnarray}
Thus,
$\cZ(\cS; \setkap)$ corresponds to
a weighted sum over all true, simply-branched, and arbitrarily pinched
$r$-sheeted covers of $\cS$,
and over the twist family for each of these.
We will use this fact to give a geometric interpretation
for Wilson loop expectation values.

\subtopic{Partition functions on surfaces without boundaries}

Partition functions on surfaces without boundary (for gauge groups
$\sun$, $\son$, and $\spn$) have been given a direct string
interpretation.\cite{gross,minahan,gt,twist,nrsstring,ram}
These partition functions
can also be obtained by cutting such a surface into pieces and
combining the partition functions  of the component surfaces with
boundary,\cite{wit} as was done in section {\it 2.b}.
The string version of this gluing process
involves sewing covers of the component surfaces together.
Since the same combinatorial factors which appear in the string picture
of Wilson loops also appear in this simpler context,
it is an instructive prelude.

Consider two surfaces $\cS_1$ and $\cS_2$
each with a single boundary component.
The partition function of the closed
surface $\cS$ obtained by gluing these
surfaces together along their boundaries is
\begin{eqnarray}
\int   \d U ~Z(\cS_1; U) Z(\cS_2; U^{-1})
& =&
\sum_{r} \sum_{ \kap \in \cK_{r}}
C_\kap \cZ(\cS_1; \kap) \cZ(\cS_2; \kap) .
\label{closedsurface}
\end{eqnarray}
The partition function on $\cS$, however,
corresponds to a weighted sum over
covers\cite{gross,minahan,gt,twist,nrsstring,ram} of $\cS$.
Equation (\ref{closedsurface})
therefore tells us how to obtain the covers of $\cS$
by sewing together
covers of $\cS_1$ and $\cS_2$ with boundary conditions
specified by the same conjugacy class $\kap$.

The factor $C_\kap$ in (\ref{closedsurface}) has two meanings:
first, it is the number of ways of sewing two
$r$-sheeted covers of a circle together,
by connecting sheets on either side at different heights
(\ie, with different labels)
in all locally distinct ways.\cite{twist}
(Since we can only join an $l_i$-fold connected component on one side with
an $l_i$-fold connected component on the other,
this sewing can only be performed if the set of connected components
over the $\cS_1$ boundary
is identical to the set of connected components
over the $\cS_2$ boundary.)
The sewing connects a sheet of a given component at a given height
to the sheets at each of the heights of the connected component
to which it is sewn on the other side
(\ie, a component with labels $(1,2, \dots, l_i)$
on one side should be sewn according to the
$l_i$ permutations $(2,3,\ldots, l_i, 1)$, {\it etc.}, on the other).
The second meaning stems from the fact that we weight each cover with
the fraction $1/|S_\nu|$.
Since some of the covers without boundary
formed by the sewing prescription are not
to be counted as distinct, the factor $C_\kap = r!/|\kap |$
overcounts in just the right way to adjust the sum over the
product of weights $1/|S_\nu|$ from the covers with boundary
so that the covers of the closed surface are correctly weighted.

Now consider a single surface $\cS$ with two boundaries.
The partition function of the surface $\cS$
formed by gluing these two boundaries
together along an orientable curve\footnote{One can also sew
the two together by antipodes, in which case one obtains a
Klein bottle insertion.} is
\eq
   \int \d U ~  Z(\cS;U,U^{-1}) = \sum_r \sum_{\kap\in \cK_r}
C_\kap  \cZ(\cS; \kap, \kap)
\en
Finally, consider a surface $\cS$ with a single boundary.
The partition function of the nonorientable surface formed by
gluing this boundary to itself by identifying antipodes is
\eq
         \int \d U ~ Z(\cS; U^2) =  \sum_r
       \sum_{\kap\in \cK_r} \eta^{r} M_\kap \cZ(\cS; \kap)
\en
where
\eq
    M_\kap = \sum_{\RinY} \delta_{R,\overline{R}} \hchi_R(\kap)
\en
plays the same role as $C_\kap$.

\subtopic{Wilson loops in string theory}

The natural gauge-invariant observables
from the perspective of string theory are not Wilson loops $W_R$
associated with some representation $R$ of a gauge group,
but rather their symmetric-group transforms,
\eq
W_\kap  = \sum_{\RinY}   \hchi_R (\kap)  W_R ,  \quad \kap \in \cK_r
\en
labeled by a conjugacy class $\kap$ of $S_r$.
The Wilson loop for the single class of the
trivial permutation group $S_1$
coincides with the usual Wilson loop taken in the fundamental representation,
$W_\one$.

The expectation value of $W_\kap$
in $\son$ and $\spn$ Yang-Mills theory on a closed surface $\cS$
has a simple geometrical intepretation:
namely, it is a weighted sum over maps from surfaces with boundaries to $\cS$,
where the boundaries are mapped to the Wilson loop.
Suppose the Wilson loop cuts the surface $\cS$ into two disjoint
surfaces $\cS_1$ and $\cS_2$,
and let the conjugacy class $\kap$ have $c$ cycles,
with cycle lengths $l_i$ ($i=1, \ldots, c$).
The expectation value of  $W_\kap$ is given by
a weighted sum over the set of maps described in sections
{\it 2} and {\it 3}
from (orientable or nonorientable)
surfaces with $c$ boundary components $C_i$
to the closed surface $\cS$,
where each of the curves $C_i$ is mapped
with degree $l_i$ to the Wilson loop on $\cS$
(\ie, traversing the curve $C_i$ once,
one circles the Wilson loop $l_i$ times).
The weights assigned to each of the maps are
the same as those for the partition function,
as described in section {\it 3}, with one difference.
The topological
factor $1/|S_\nu|$ should be computed so that any sheet of the cover
that ends on the Wilson loop is considered to carry a distinct label.
(For example, a completely disconnected $r$-fold cover of a Wilson
loop on the sphere should be counted with weight $1$, not $1/r!$.)

The maps from open worldsheets $\cW$ to $\cS$
divide into disjoint sets according to
which surface ($\cS_1$ or $\cS_2$)
the neighborhood of each boundary component $C_i$ covers.
Specifically, one considers all possible divisions
of the cycles of $\kap$
into two sets $\kap_1$ and $\kap_2$ (either of which could be empty);
the boundary components corresponding to cycles in $\kap_1$ or $\kap_2$
have neighborhoods which map to $\cS_1$ and $\cS_2$ respectively.

To construct the set of maps just described,
one glues together appropriate covers of $\cS_1$ and $\cS_2$,
as in the construction of the partition function on a closed surface
detailed in section {\it 4.b}.
Consider the maps corresponding to the
division $\kap = \kap_1 \cdot \kap_2 \in S_{r_1} \oplus S_{r_2}$ (with
$r_1 + r_2 = r$).
Take all covers of $\cS_1$
in which the sheets cover the $\cS_1$ boundary according to
$ \kap_1 \cdot \pi $
(for any class $\pi \in \cK_{p}$ for any $p$),
and all covers of $\cS_2$ in which the sheets cover the
$\cS_2$ boundary according to
$\kap_2 \cdot \pi $,
and join together just
the components of the boundaries involving the $p$ sheets of $\pi$
on each side in exactly the manner described in the previous section.
This gives a covering surface of $\cS$
whose boundary is mapped to the Wilson loop
with the correct boundary conditions.
By summing over all conjugacy classes $\pi \in \cK_{p}$
and all $p \ge 0 $,
one obtains a correctly weighted sum over all covers that enter.
Next, one sums over all possible divisions of
the cycles of $\kap$ into $\kap_1$ and $\kap_2$
to obtain all maps corresponding to $\kap$.

To establish the validity of this interpretation,
we compute the Wilson loop expectation value
in $\son$ or $\spn$ Yang-Mills theory
\begin{eqnarray}
W_\kap
& = &
\int \d U ~Z(\cS_1; U) \Upkap (U) Z(\cS_2; U^{-1})
\nonumber\\
& = &
\sum_{\lambda,\mu}  \cZ(\cS_1; \lambda) \cZ(\cS_2; \mu)
\int \d U ~\Up_{\lambda} (U) \Upkap (U)  \Up_{\mu} (U^{-1}) .
\label{firststep}
\end{eqnarray}
To evaluate this,
we need an expression for a product of $\Upkap (U)$'s.
Using eqs.~(\ref{schurdef}) and (\ref{charproduct}), we find
\begin{eqnarray}
\Up_\lambda (U) \Up_\kap (U)
 = \sum_{R,S,T}
\hchi_R (\lambda)  \hchi_S (\kap)  \NG{R}{S}{T} \gchi_T (U).
\label{secondstep}
\end{eqnarray}
If $N$ is large, the $\sun$ tensor multiplicities
equal the Littlewood-Richardson coefficients
\eq
 N_{\sun} (R,S;T) = L (R, S; T) ,
\en
for which $r(R) + r(S) = r(T)$.
Using\cite{symmbook}
\eq
\hchi_R (\kap_1 \cdot \kap_2)
= \sum_{R_1 \in Y_{r_1} } \sum_{R_2 \in Y_{r_2} }
L(R_1, R_2; R) \hchi_{R_1} (\kap_1) \hchi_{R_2} (\kap_2) ,
\label{symident}
\en
we obtain
\eq
\Up_\lambda (U) \Up_\kap (U)
= \Up_{\lambda \cdot \kap} (U),
\qquad {\rm for~}G = \sun,
\en
where $\lambda \cdot \kap$
is a class of $S_{r_\lambda} \oplus S_{r_\kap}$.
Similarly, given that $N$ is large enough,
the $\son$ and $\spn$ tensor
multiplicities satisfy\cite{tensorprod}
\eq
\gchi_R (U) \gchi_S(U) = \sum_T \sum_C L(R/C,S/C;T)  \gchi_T (U),
\en
in which
\eq
R/C \equiv \sum_D L(C,D;R) D,
\en
so that the $\son$ and $\spn$ tensor multiplicities
can also be written in terms of Littlewood-Richardson
coefficients\cite{tensorprod}
\eq
N_{\son}  (R,S;T) = N_{\spn}  (R,S;T)
=  \sum_{ R_1, R_2, R_3 }
L(R_1,R_3;R) L(R_1,R_2;S) L(R_2,R_3;T).
\label{sontensor}
\en
As a result, we only retain $r(R) + r(S) = r(T) \; {\rm mod}\; 2$.
{}From (\ref{symident}), we have
\eq
L(R_1, R_2; R)
=
 \sum_{\kap_1 \in \cK_{r_1} } \sum_{\kap_2 \in \cK_{r_2} }
 C_{\kap_1}^{-1} C_{\kap_2}^{-1}
\hchi_{R_1} (\kap_1) \hchi_{R_2} (\kap_2) \hchi_R (\kap_1 \cdot \kap_2) ,
\en
so that, since  $C_{\kap_1 \cdot \kap_2} = C_{\kap_1} C_{\kap_2}  $,
\eq
\sum_{R \in Y_{r_1+r_2} }
\hchi_R (\kap) L(R_1, R_2; R)
=
 \sum_{\kap_1 \in \cK_{r_1} } \sum_{\kap_2 \in \cK_{r_2} }
\hchi_{R_1} (\kap_1) \hchi_{R_2} (\kap_2)
\delta_{\kap, \kap_1 \cdot \kap_2 }.
\label{symidenttwo}
\en
Inserting eq.~(\ref{sontensor}) into (\ref{secondstep}),
and employing eqs.~(\ref{symident}) and (\ref{symidenttwo}),
we find after some simplification
\eq
\Up_\lambda (U) \Up_\kap (U) =
\sum_{\kap_1,\kap_2,\kap_3} C_{\kap_1}
\delta_{\lambda, \kap_1 \cdot \kap_3} \delta_{\kap, \kap_1 \cdot \kap_2}
\Up_{\kap_2 \cdot \kap_3}(U),
	\quad {\rm for~}G = \son{\rm~and~}\spn.
\en
Inserting this into eq.~(\ref{firststep}), and using
orthogonality (\ref{schurorthonormality}), we obtain
\begin{eqnarray}
  W_\kap & = & \sum_{\kap_1,\kap_2,\kap_3} C_{\kap_1} C_{\kap_2} C_{\kap_3}
            \cZ(\cS_1; \kap_1\cdot \kap_3) \cZ(\cS_2; \kap_2 \cdot \kap_3)
          \delta_{\kap,\kap_1 \cdot \kap_2} \nonumber\\
  & = & C_\kap \sum_{\kap_1,\kap_2}
          \delta_{\kap,\kap_1 \cdot \kap_2}
           \sum_{p} \sum_{\pi\in \cK_{p}}
          C_{\pi}  \cZ(\cS_1; \kap_1\cdot \pi)
              \cZ(\cS_2; \kap_2 \cdot \pi).
\label{keyresult}
\end{eqnarray}
This result for the Wilson loop expectation value has the very simple
geometric interpretation given above;
namely, it is a weighted sum over surfaces to $\cS$.
Those surfaces can have sheets over $\cS$
on both sides of the Wilson loop,
with some sheets from one side terminating at the Wilson loop
(corresponding to permutation  $\kap_1$),
some sheets from the other side terminating at the loop
(corresponding to permutation  $\kap_2$),
and the remaining sheets on both sides joining together at the loop
(corresponding to permutation  $\pi$).
Thus we have confirmed the string interpretation
of the Wilson loop expectation values $W_\kap$ in $\son$
and $\spn$ Yang-Mills theory.
The only difference between the string theory description of $\son$
and that of $\spn$
is the sign $\sign$ associated with cross-caps
in the weight assigned to each string theory map.
Compared to the analogous result for a chiral sector of  $\sun$ Yang-Mills
theory,
\eq
    W^{\rm ch}_{\kap}
= C_\kap \sum_{p} \sum_{\pi\in \cK_{p}}
  C_\pi \cZ(\cS_1; \pi) \cZ(\cS_2; \kap \cdot \pi) ,
\label{chiralresult}
\en
which has sheets that end over the Wilson loop
extending to only one side,
we see that in the string theory corresponding to
$\son$ and $\spn$ Yang-Mills theory,
the covers extend to both sides
(a feature that only appears in the coupled $\sun$ sectors).

For homologically nontrivial orientable Wilson loops,
the $\son$ and $\spn$ expectation value is
\eq
   W_\kap = C_\kap
\sum_{\kap_1,\kap_2} \delta_{r_1,r_2} \delta_{\kap,\kap_1 \cdot \kap_2}
\sum_{p} \sum_{\pi\in \cK_{p}}
C_{\pi}  \cZ(\cS; \kap_1\cdot \pi, \kap_2 \cdot \pi).
\en
This vanishes unless the number of sheets $r_\kap$
terminating at the Wilson loop is even,
which is the string analog of the gauge theory
conservation law described in section {\it 2.c}.
The result for a chiral $\sun$ sector along the same curve,
\eq
    W^{\rm ch}_{\kap}
 = C_\kap \sum_{\pi}
 \cZ(\cS; \pi, \pi \cdot \kap) C_{\pi} =
  \delta_{r_\kap, 0} \sum_p \sum_{\pi\in \cK_p} C_\pi \cZ(\cS; \pi, \pi),
\en
vanishes unless $\kap$ is trivial.
Finally, for homologically nontrivial, nonorientable Wilson loops
(for $\son$, $\spn$, or chiral $\sun$),
the expectation value is
\eq
W_\kap
= \int \d U ~ \cZ(\cS; U^2) \Up_\kap(U)
=  \sum_{p} \sum_{\pi\in \cK_{p}}
               \cZ(\cS; \pi) M_{\kap,\pi}
\en
with the integer
\eq
 M_{\kap,\pi} = \sum_{R\in Y_{r_\kap}} \sum_{P\in Y_p}
({N^+_{PP}}^R - {N^-_{PP}}^R) \hchi_{R}(\kap) \hchi_{P}(\pi)
\en
playing the same role for antipodal sewing
that $C_\kap C_\pi$ does for ordinary sewing.

\topic{Illustrations}

Although the covers that contribute to the
expectation value of a general nonintersecting Wilson loop
in $\son$ and $\spn$ Yang-Mills theory
have been fully described in the last section,
a few explicit examples may help to illuminate the general formulae
and make transparent the equivalence of the string and gauge theory results.
The contributing covers with $r$ sheets correspond directly to the
gauge-picture terms labeled by Young tableaux with $r$ cells.
Therefore, we can display the equivalence by
forming the weighted sum over all $r$-sheeted
covers described in section {\it 3}
with $r$ ranging up to a fixed number of sheets,
and by then comparing this with the corresponding terms
of the gauge theory expansion given in section {\it 2}.

\subtopic{Wilson loop on a sphere}

Consider the Wilson loop $W_\kap$ on the sphere,
where $\kap$ is the single conjugacy class $(1)$
of the trivial permutation group $S_1$;
$W_{(1)}$ coincides with  $W_\one$,
the Wilson loop in the fundamental representation of the gauge group.
The Wilson loop divides the sphere into two disks,
$D_1$ and $D_2$,
each with unit Euler characteristic,
and with (dimensionless) areas $\cA_1$ and $\cA_2$ respectively.
To compute the expectation value $W_{(1)}$,
we must consider all maps from (connected and disconnected)
surfaces with a single boundary component to the sphere,
with the boundary mapped one-to-one to the Wilson loop.
One such map is that from a disk to $D_1$,
whose weight is $ N \exp ( - \cA_1/2 ) $.
We must also consider maps from a disk
with an arbitrary number of cross-cap insertions
to $D_1$;
summing over all possible cross-cap insertions, we obtain
\eq
N   \exp \left( -{\cA_1 \over 2} \right)
\sum_{\ncc=0}^\infty  {1\over \ncc!} \left( \sign \cA_1 \over 2N \right)^\ncc
= N   \exp \left( -{\cA_1 \over 2} + {\sign \cA_1 \over 2N}   \right)
\equiv  N E (\cA_1).
\en
There are equivalent maps to $D_2$,
whose weight is obtained by replacing $\cA_1$ with $\cA_2$.
Because these are degree-one maps,
there are no branch points
and the twist operator $\Omega_1$ is the identity.

To determine the next maps that contribute,
take the double cover of $D_1$
consisting of two disjoint disks,
and join one of the sheets of this cover
to the single cover of $D_2$.
The resulting surface is the disconnected sum of a sphere and a disk,
which has Euler characteristic three,
with the boundary of the disk mapped to the Wilson loop.
Including possible cross-cap insertions,
these maps give the contribution
\eq
N^3   E (2 \cA_1 + \cA_2 ) .
\en

Since there are two sheets over $D_1$,
the twist operator is nontrivial
and two-sheeted simply-branched covers of $D_1$ must also be included.
{}From eq.~(\ref{omegadef}),
the twist operator is
\eq
\Om_2 = \left( 1 - {1\over N^2} \right) {\bf 1}
+ \left( {1 \over N}  - {\sign \over N^2} \right) {\bf P},
 \en
where $\bfid = (1)(2)$ and ${\bf P} = (12)$.
Ramgoolam\cite{ram} has given a geometric interpretation of
the subleading terms of $\son$ and $\spn$ twist operators in
terms of insertions of infinitesimal tubes and cross-caps located
at the twist points.
In the case of $\Om_2$,
the leading term $\bfid$ is the identity permutation,
while the subleading term $-\bfid/N^2$ corresponds to the
insertion of a tube connecting the two sheets at the twist point.
${\bf P}/N$ corresponds to the insertion of a simple branch point,
either alone or with a cross-cap insertion (with weight $-\eta/N$).
We may also have arbitrary numbers of simple branch points
that connect the two sheets over $D_1$ and that contribute an area
dependence to the weight
(unlike the simple branch point from the twist operator).
The first term of the twist operator acts trivially on the sheets,
so that any accompanying simple branch points must
come in pairs\cite{nrstwist} connected by double arcs. These
covers yield the contribution
\eq
\Beven
= \left( 1 - {1 \over N^2} \right)
\sum_{i=0}^\infty {1 \over (2i)!} \left(  - \cA_1 \over N \right)^{2i}
= \left( 1 - {1 \over N^2} \right)
\cosh \left(- \cA_1 \over N \right).
\en
The second term of the twist operator
must be acompanied by an odd number of simple branch
points, with one of the area-dependent simple branch points connected
by a double arc to the simple branch point from the twist operator.
These covers yield the contribution
\eq
\Bodd
= \left( {1 \over N} - {\sign \over N^2} \right)
\sum_{i=0}^\infty {1 \over (2i+1)!} \left(  - \cA_1 \over N \right)^{2i+1}
= \left( {1 \over N} - {\sign \over N^2} \right)
\sinh \left( - \cA_1 \over N \right).
\en
Similarly, there are maps where the double cover is over $D_2$.
Adding up all the contributions obtained so far, we have
\eq
W_{(1)} =
N E (\cA_1) + N^3  (\Beven + \Bodd)  E (2 \cA_1 + \cA_2 )
+ ( \cA_1 \leftrightarrow \cA_2) + \cdots  .
\en
Using the expression for the Casimirs (\ref{things})
and the dimensions of the representations (see ref.~\dcite{nrsstring}),
we may rewrite this as
\begin{eqnarray}
W_\one
& = &  (\dim \bone)  (\dim \bfid)
  \exp \left( - \lam A_1 c_{\;\one} \over 2N \right)
+ (\dim \bone)   (\dim \btwo)
  \exp \left( - \lam
\left[ A_1 c_{\;\one}  + A_2 c_{\;\two} \right]
 \over 2N \right)
\nonumber \\
& +  & (\dim \bone) (\dim \boneone)
 \exp \left( - \lam
\left[ A_1 c_{\; \one}  + A_2 c_{\; {\raise 1mm \hbox{\oneone}}}
                     \right]  \over 2N \right)
+ ( A_1 \leftrightarrow A_2 ) + \cdots,
\end{eqnarray}
which are the first few terms in the gauge theory expression
for the Wilson loop $W_\one$  (\ref{homologtrivial}).

\subtopic{Wilson loop on $RP_2$}

Next, consider a homologically trivial
Wilson loop on the real projective plane, $RP_2$,
which divides it into a disk $D$ of area $\cA_1$
and a M\"obius strip $M$ of area $\cA_2$.
The contribution of maps
from a disk (with cross-cap insertions) to $D$ is
$ N E (\cA_1) $.
Maps from a M\"obius strip (with cross-cap insertions) to $M$
give the contribution
$ \sign E (\cA_2) $.
One of the sheets of the double cover of $D$
can be joined to a single cover of $M$
to give the disconnected sum of a disk and a real projective plane;
maps from this surface contribute
$ \sign N^2 (\Beven + \Bodd) E ( 2 \cA_1 + \cA_2 )  $,
allowing for maps with simple branch points and twist points
in the double cover of $D$ .
Finally, there are two possible double covers of $M$:
(1) a disconnected surface consisting of two M\"obius strips, and
(2) a cylinder.
The first can be glued to a disk covering $D$ to
give a disconnected sum of a M\"obius strip and
a real projective plane,
whereas the second can be glued to a disk covering $D$
to yield a larger disk of area $2\cA_1 + \cA_2$.
Maps from each of these contribute
$ N E ( 2 \cA_2 + \cA_1 ) \cosh ( -\cA_2 / N  )  $,
including the possibility of maps with an even
number of simple branch points.
(Since $M$ has vanishing Euler characteristic,
there are no twist points.)
The sum of these contributions is
\begin{eqnarray}
W_{(1)}
 =
   N E (\cA_1)
&+&\sign E (\cA_2)
 + \sign N^2 (\Beven + \Bodd) E ( 2 \cA_1 + \cA_2 )
\nonumber \\
&+&  2N E ( 2 \cA_2 + \cA_1 ) \cosh \left(-\cA_2 \over N  \right)
+ \cdots,
\end{eqnarray}
agreeing to this order with the gauge theory result
\begin{eqnarray}
W_\one
& = &
(\dim \bone) \sign_{{\bf 1}}
  \exp \left( - \lam A_1 c_{\; \one} \over 2N \right)
+ (\dim \bfid) \sign_{\; \one}
  \exp \left( - \lam A_2 c_{\; \one} \over 2N \right)
\nonumber \\
&+&   (\dim \bone)   \sign_{\; \two}
  \exp \left( - \lam \left[ A_1 c_{\; \one}
                 + A_2 c_{\; \two} \right] \over 2N \right)
+ (\dim \btwo) \sign_{\one}
  \exp \left( - \lam \left[ A_1 c_{\; \two}  +
            A_2 c_{\; \one} \right] \over 2N \right)
 \nonumber \\
&+&   (\dim \bone) \sign_\oneone
\exp \left( - \lam \left[ A_1 c_{\; \one}
          + A_2 c_{\; {\raise 0.5mm \hbox{\oneone}}} \right]  \over 2N \right)
 + (\dim \boneone) \sign_{\one}
\exp \left( - \lam \left[ A_1 c_{\; {\raise 0.5mm \hbox{\oneone}}}
                           + A_2 c_{\; \one} \right]
\over 2N \right) + \cdots.
\nonumber \\
\end{eqnarray}

\subtopic{Homologically nontrivial Wilson loops on a torus}

Finally, we consider homologically nontrivial Wilson loops on a torus.
Since the expectation value of a
single homologically nontrivial Wilson loop $W_\kap$ vanishes
automatically unless $r_\kap$ is even
(the number of sheets cannot change as you cross the loop
since it borders the same region on both sides),
$W_{(1)}$ vanishes.

We consider a pair of Wilson loops,
which cut the torus into two cylinders
$C_1$ and $C_2$,
with areas $\cA_1$ and $\cA_2$ respectively.
Its expectation value corresponds to a weighted
sum over all maps from (connected and disconnected)
surfaces with two boundary components
that are mapped to the two Wilson loops.
First, there are maps from a cylinder
to either $C_1$ or $C_2$,
with an arbitrary number of cross-cap insertions,
contributing
$  E (\cA_1) +  E (\cA_2)  $.
Next,
consider the double cover of $C_1$ consisting of
a pair of disjoint cylinders (for which $1/|S_\nu| = 1/2$),
and the single cover of $C_2$.
There are four possible ways to join
these covers but only two distinct sewn covers:
one yields the disconnected sum of a torus and cylinder,
and the other a long cylinder of area $2\cA_1 + \cA_2$
(each are produced twice, cancelling the topological factor $1/2$).
The number of simple branch points in the double cover
of $C_1$ must be even,
because there are no twist points,
so each of these surfaces contributes
$   E ( 2\cA_1 + \cA_2 ) \cosh (- \cA_1 / N )  $
The double cover could also map to $C_2$, so
that the final sum is given by
\eq
  E (\cA_1)
+ 2  E ( 2\cA_1 + \cA_2 ) \cosh (- \cA_1 / N )
+ (\cA_1 \leftrightarrow \cA_2) + \cdots,
\en
which agrees with the gauge theory result
\begin{eqnarray}
  \exp \left( - \lam A_1 c_{\;\one} \over 2N \right)
& +& \exp \left( - \lam
\left[ A_1 c_{\;\one}  + A_2 c_{\;\two} \right]
 \over 2N \right)
\nonumber \\
& +  & \exp \left( - \lam
\left[ A_1 c_{\; \one}  + A_2 c_{\; {\raise 0.1mm \hbox{\oneone}}}
                     \right]  \over 2N \right)
+ ( A_1 \leftrightarrow A_2 ) + \cdots.
\end{eqnarray}

\end{document}